\documentclass[conference]{IEEEtran}
\IEEEoverridecommandlockouts
\usepackage{graphicx}
\usepackage{amsmath,amsfonts,amssymb,amsthm}
\usepackage{balance}
\usepackage{cite}
\usepackage{fixmath}
\usepackage{gensymb}
\usepackage[center]{caption}
\usepackage{subcaption}
\usepackage{multirow}
\newtheorem{lem}{Lemma}

\usepackage[english]{babel}
\usepackage{cite}
\usepackage{epstopdf}
\usepackage{amsmath,amssymb,amsthm,mathrsfs,amsfonts,dsfont}
\usepackage{algorithmic}
\usepackage{graphicx}
\usepackage{graphics}
\usepackage{color}
\usepackage{textcomp}
\usepackage{balance}
\usepackage[top=0.75in,bottom=1in,left=0.675in,right=0.675in]{geometry}

\newcommand{\site}{\boldsymbol{s}}
\def\BibTeX{{\rm B\kern-.05em{\sc i\kern-.025em b}\kern-.08em
		T\kern-.1667em\lower.7ex\hbox{E}\kern-.125emX}}
\begin{document}
	
	\title{A 3D Beamforming Scheme Based on The Spatial Distribution of User Locations}

	\author{\IEEEauthorblockN{Jalal Rachad$~^{*+}$, Ridha Nasri$~^*$, Laurent Decreusefond$~^+$}
		\IEEEauthorblockA{$~^*$Orange Labs: Direction of Green transformation, Data knowledge, traffic and resources Modeling,\\ 40-48 avenue de la Republique
			92320 Chatillon, France\\ 
			$~^+$LTCI, Telecom Paris, Institut Polytechnique de Paris,\\
			 75013, Paris, France\\
			Email:$^{*}$jalal.rachad@ieee.org,  $~^*$ridha.nasri@orange.com,
      $^+$Laurent.Decreusefond@mines-telecom.fr}
}		
		
%
%
%
	\maketitle
	\begin{abstract}
  Multi-antenna technologies such as massive Multiple-Input Multiple-Output (massive MIMO) and beamforming are key features to enhance performance, in terms of capacity and coverage, by using a large number of antennas intelligently. With the upcoming 5G New Radio (NR), FD-MIMO (Full Dimension MIMO) will play a major key role. FD-MIMO consists in arranging a large number of antennas in a 2D array, which enables to use 3D beamforming i.e., beamforming in both horizontal and vertical dimensions. The present paper provides a 3D beamforming model where beam steering depends on the random spatial distribution of users. We attempt to derive some analytical results regarding the probability distribution of antenna beamforming radiation pattern. Also, through system level simulations, we show how 3D beamforming can reduce interference impact, compared to the traditional 2D beamforming, and enhances system performance in terms of the coverage probability and users throughput.    

	\end{abstract}
	
	
	\section{Introduction}
 As the amount of wireless data traffic is increasing continuously, 5G networks need to support this proliferation through upgrading features in order to ensure a strong compatibility with the upcoming mobile network generation services. Massive Multiple antenna technologies including 3D beamforming have drawn recently the attention of telecommunication actors and research community. Actually, Beamforming consists in forming a signal beam between the transmitter and the receiver by using an array of antennas. It enhances the signal strength at the receiver and minimize interference level so that high average data rate and high spectral efficiency can be achieved. Most existing base stations (BSs) are equipped with directional antennas that provides radiation patterns in the horizontal dimension, considering only the azimuth angle, and having a fixed vertical pattern and downtilt. Recently, there has been a trend to consider also the impact of the elevation and investigate how antennas downtilt influence performance. It has been shown that the combination of the horizontal and the vertical dimensions can further improve the signal strength at the receiver location, enhances data rates and minimize interference in neighboring cells \cite{gandhi2014significant}.\\
  
  MIMO and beamforming have been extensively studied in literature. For instance in \cite{bai2015coverage}, coverage and rate in mmWaves network have been analyzed in an outdoor environment. Authors considered directional antennas, approximated by sectored antenna model, to perform directional beamforming in BSs and user equipments (UEs). In \cite{koppenborg20123d} authors investigated the impact of adapting the vertical pattern of antennas to user locations through conducted lab and field trial measurements. They showed that, either in outdoor or indoor environments with Line Of Sight (LOS) and Non Line Of Sight (NLOS) conditions, adapting the vertical pattern improves performance. Interference restraint using dynamic vertical beam steering has been studied in \cite{halbauer2012interference}. In \cite{yang2018optimal} authors have analyzed the impact of antenna radiation pattern and the downtilt on the performance of a Downlink (DL) cellular network in terms of the coverage probability and the area spectral efficiency (ASE). For the analytical approach, they have used tools from stochastic geometry to model the spatial distribution of nodes. Also, they have shown that there exists an optimal antenna downtilt that depends on the LOS and NLOS conditions. This optimal downtilt maximizes the coverage probability and improves the ASE. Similarly in \cite{seifi2012impact}, the effect of antenna downtilt in a MIMO system has been studied through system level simulations. Authors used the 3rd Generation Partnership Project (3GPP) antenna radiation pattern model to show that the maximal cell throughput depends on the antenna downtilt. Additionally, the network geometry and the spatial distribution of users are important factors to consider when looking for system performance analysis. Most considered models that can be found in literature are the hexagonal model \cite{nasri2016analytical} and \cite{rachad2018interference}, and random models based on spatial point process \cite{andrews2011tractable} and \cite{blaszczyszyn2016spatial}. Analytical formulas are always considered hard to derive when BSs are considered equipped with sectorized antennas, especially in the case of random models, since interference depends on the antenna radiation pattern in each sector. In \cite{nasri2016analytical}, an explicit expression of Interference to Signal Ratio (ISR) has been derived in a regular hexagonal tri-sectorized network considering the horizontal antenna radiation pattern. In \cite{blaszczyszyn2016spatial} authors have analyzed performance of sectorized random network in terms of Signal to Interference plus Noise Ratio (SINR). They have shown that the distribution of SINR is insensitive to the antenna azimuth's distribution.\\
  
  The main contribution of this paper is to propose a new analytical approach to model 3D beamforming in a regular tri-sectorized hexagonal network. We characterize at first the antenna beamforming radiation pattern, in both horizontal and vertical dimensions, considering the random locations of users in the cells. Also, we prove that $ISR$ is an almost sure convergent series of independent random variables and we show how to derive the mathematical expectation of $ISR$ in each location of a mobile in a typical serving cell. This metric is useful for link budget tools in which the expression of the average perceived interference is required in each position. Then through system level simulations, we analyze performance in terms of the coverage probability and users throughput. We compare the proposed 3D beamforming model with 2D beamforming in terms of coverage probability enhancement and interference reduction.\\
  
  The rest of the paper is structured as follows. In section II, we present our network model as well as the 3D beamforming model. Section III provides a characterization of antennas radiation patterns under the assumptions of the 3D beamforming model. A brief characterization of interference is given in section IV. Numerical results are provided in section V. Section VI concludes the paper.

\section{System model and notations}
\subsection{Network model}

We consider a regular tri-sectorized hexagonal network denoted by $\Lambda$ with an infinite number of sites $\site$  having an inter-site distance denoted by $\delta$. For each site $\site$ $\in$ $\Lambda$, there exists a unique (u,v) $\in$ $\mathbb{Z}^2$ such that $\site = \delta( u + ve^{i\frac{\pi}{3}})$, we denote by $\site_0$ the name of the serving cell located at the origin of $\mathbb{R}^2$. Unlike regular hexagonal network with omni-directional antennas, BSs of sites are located at the corner of the hexagons. All BSs have the same height $l_b$,  transmit with the same power level $P$ and assumed to have directional antennas covering each one a hexagonal sector identified by $c \in \{1,2,3 \}$. The azimuths of antennas $\vartheta_c$ in which the radiation is at its maximum is taken relative to the real axis and it is given by 

\begin{equation}
\vartheta_c=\frac{\pi}{3}(2c-1),
\label{azimuths}
\end{equation}
so that the azimuth of the first sector of each site has an angle of $\frac{\pi}{3}$ with the real axis relatively to the position $\site$.\\

The location of a mobile served by the first sector ($c=1$) of $\site_{0}$ is denoted by $m$ such that $m = re^{i\theta}$ where (r, $\theta$) are the polar coordinates in the complex plane. We denote also by $n_{\site,c}$ the geographical location of a mobile served by a sector $c$ of a site $\site$ $\in$ $\Lambda^{*}$ in the plane, where $\Lambda^{*}$ is the lattice $\Lambda$ without the serving cell $\site_0$. Locations $n_{\site,c}$ are written in the complex plane by $n_{\site,c} = \site + r_{\site,c} e^{i\theta_{\site,c}}$, where $r_{\site,c}$ and $\theta_{\site,c}$ represent respectively the distance and the angle (complex argument) between $n_{\site,c}$ and $\site$.\\

\subsection{Beamforming model}

As we mentioned previously, BSs are equipped with directional antennas with sectorized gain pattern. At each TTI (Time Transmit Interval), we assume that there is at most one user served by a sector $c$ of site $\site$ using beamforming radiation pattern. This can be modeled by a Bernoulli RV $\rho_{\site,c}$ such that $\mathbb{P}(\rho_{\site,c}=1)=\eta$ and $\mathbb{P}(\rho_{\site,c}=0)=1-\eta$. $\eta$ represents the percentage of the occupied resources or the average load over the interfering sites $\site$.\\

 Furthermore, we assume that each antenna has a directional radiation that can be described by two planar patterns: the horizontal and the vertical one denoted respectively by $H$ and $V$. We define the antenna radiation for each pattern by a $2\pi$-periodic function $U$ such that its restriction on  $ (-\pi, \pi] \rightarrow [0,1]$ has the following properties \\
\begin{description}
	\item[$\bullet$] $U(\alpha)=U(-\alpha)$ for all $\alpha \in (-\pi,\pi]$,\\
	
	\item[$\bullet$] $U(\alpha)=0$ if $ \frac{\pi}{2} \leq|\alpha|\leq \pi $,\\
	
	\item[$\bullet$] $U$ is decreasing on $[0,\frac{\pi}{2}]$ and increasing on $[\frac{-\pi}{2},0]$.\\
\end{description}    

Numerous antenna radiation pattern models can be found in literature such as the 3GPP model, Mogensen model \cite{mogensen1997preliminary} and real antenna patterns provided by constructors (Kathrein antennas...). For reasons of tractability, we adopt in the reminder of this analysis Mogensen model to describe the antenna horizontal and vertical radiation patterns in the linear scale as follows

	\begin{align}
	&H(\alpha)= [\cos(\alpha)]^{-2w_h} \\
	&V(\phi)=[\cos(\phi)]^{-2w_v},
	\label{Mogensen}
	\end{align} 
with $w_h=\frac{ln(2)}{\ln(\cos(\frac{\theta_{h3dB}}{2})^2)}$ and $w_v=\frac{ln(2)}{\ln(\cos(\frac{\theta_{v3dB}}{2})^2)}$. $\theta_{h3dB}$ and $\theta_{v3dB}$ are respectively the horizontal and the vertical half power beam widths.\\ 

This model can be adapted to beamforming radiation pattern by varying the beam widths ($\theta_{h3dB}$ and $\theta_{v3dB}$ ). Hence, the antenna radiation pattern received in a mobile location $m$ from an interfering site $\site$ is defined by

\begin{equation}
G_{\site}(m)=\sum_{c=1}^{3} \rho_{\site,c} H(\alpha_{\site,c})V(\phi_{\site,c}),
\label{gain}
\end{equation} 
with $\alpha_{\site,c}$ is the angle between the mobile $m$ orientation and the beam axis directed to a mobile $n_{\site,c}$ in the horizontal plane and $\phi_{\site,c}$ is the angle between the beam direction in the vertical plane and the mobile $m$. The angle $\alpha_{\site,c}$ can be expressed, based on the complex geometry, as

\begin{equation}
\alpha_{\site,c}=  \psi(m,\site) - \theta_{\site,c}, 
\label{alpha}
\end{equation} 
where $ \psi(m,\site)=arg(m-\site)$ and $\theta_{\site,c}=\psi(n_{\site,c},\site)$ is the complex argument of $n_{\site,c}$ relatively to $\site$. Each $\theta_{\site,c}$ is assumed to be uniformly distributed in the interval $[\vartheta_c-\frac{\pi}{3},\vartheta_c+\frac{\pi}{3}]$, thus using a linear transformation of the RV $\theta_{\site,c}$, we can easily prove that the angle $\alpha_{\site,c}$ is a RV uniformly distributed in the interval $[ \psi(m,\site)-\frac{2\pi}{3}c, \psi(m,\site)+\frac{2\pi}{3}(1-c)]$.\\

Similarly for the vertical dimension,  the angle $\phi_{\site,c}$ can be expressed as
\begin{equation}
\phi_{\site,c}= atan(\frac{l_b}{|m-\site|})- \tilde{\phi}_{\site,c},
\label{phi1}
\end{equation}
with $\tilde{\phi}_{\site,c}=atan(\frac{l_b}{r_{\site,c}})$  refers to the antenna downtilt, which is variable in our case.\\

\begin{figure}[tb]
	\centering
	\includegraphics[width=7cm,height=6cm]{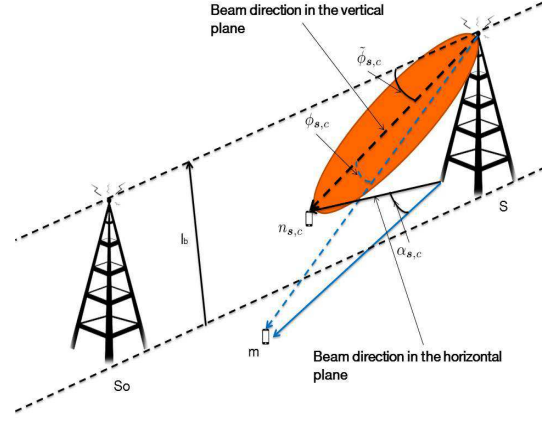}
	\caption{3D beamforming illustration}
	\label{hexagonal}
\end{figure}

 The distance $r_{\site,c}=|n_{\site,c}-\site|$ between a mobile $n_{\site,c}$ and $\site$ varies between $0$, when $n_{\site,c}$ is close to $\site$ location, and $\frac{2\delta}{3}$ when $n_{\site,c}$ is located at the edge of a sector. This distance can be characterized using the antenna radiation pattern covering a whole hexagonal sector of $\site$. Thus, $r_{\site,c}$ is varying between $0$ and $\frac{2\delta}{3}U(\theta_{\site,c}-\vartheta_c)$, with $U(\theta_{\site,c}-\vartheta_c)$ is the antenna radiation pattern of a sector (i.e., the half power beam width is equal to 65 degrees). So the mobile will be located at the far edge of a sector when the angle between $n_{\site,c}$ and $\site$ is equal to the antenna azimuth in which the radiation is at its maximum. Moreover, since mobile locations $n_{\site,c}$ are assumed to be uniformly distributed, we consider in the remainder that $r_{\site,c}$ is a RV uniformly distributed on the interval $[0,\frac{2\delta}{3}U(\theta_{\site,c}-\vartheta_c)]$. Fig. \ref{hexagonal} illustrates the 3D beamforming model.\\

\subsection{Propagation model}

To model the wireless channel, we consider the standard power-law path loss model based on the distance between a mobile $m$ and a BS $\site$ such that the path loss $L(\site,m)$ is given by

\begin{equation}
L(\site,m)=a|\site-m|^{2b},
\label{pathloss}
\end{equation}
with $2b$ is the path loss exponent and $a$ is a propagation factor that depends on the type of the environment (Indoor, Outdoor...).\\

In addition to the path loss, the received power by a mobile served in downlink depends on the random channel effects, especially shadowing and fast fading. Shadowing refers to the attenuation of the received signal power caused by obstacles obstructing the propagation between the transmitter and receiver. In this paper, we model the shadowing effect between a mobile location $m$ and an interfering site $\site$ by a log-normal RV $\chi_{\site}(m)=10^{\frac{Y_{\site}(m)}{10}}$ with $Y_{\site}(m)$ is a Normal RV with mean $\mathbb{E}(Y_{\site}(m))=0$ and variance $\sigma^2$. The shadowing effect between $m$ and the serving site $\site_0$ is denoted by $\chi_0$. This sequence of RVs are assumed to be independent and identically distributed for all $(\site,m)$.\\

On the other hand, fast fading random model is not considered in this paper. Its effect can be compensated through link level performing that maps the $SINR$ to the throughput ($Th$). Also, for an AWGN (Additive Gaussian Noise Channel), Shannon's formula provides the relation between $SINR$ and $Th$. Hence, the fast fading effect can be compensated by using a modified Shannon's formula to have $Th=K_1 log_2(1+K_2 SINR)$, with $K_1$ and $K_2$ are constants calibrated from practical systems \cite{mogensen2007lte}.\\

 Therefore, the received power from a BS $\site$, transmitting with a power level $P$, in a mobile location $m$ is expressed by

\begin{equation}
P_r(m,\site)=\frac{P A G_{\site}(m) \chi_{\site}(m)}{L(\site,m)},
\label{receivedpower}
\end{equation}   
with $A$ is the antenna gain and $G_{\site}(m)$ is the antenna radiation pattern from $\site$ received at the mobile location $m$.\\ 

\section{Characterization of antenna beamforming radiation pattern $G_{\site}(m)$}

It is obvious from (\ref{gain}) that $G_{\site}(m)$ is a RV that depends on the three RVs: $\alpha_{\site,c}$, $\phi_{\site,c}$ and $r_{\site,c}$. Also, one can notice that $V(\phi_{\site,c})$ is a RV that depends on the RV $H(\alpha_{\site,c})$. Therefore, to characterize $G_{\site}(m)$, we start first by deriving the probability density function (PDF) of $H(\alpha_{\site,c})$ and the PDF of $V(\phi_{\site,c})$ conditionally on $H(\alpha_{\site,c})$, denoted respectively by $f_{H_{\site,c}}(h)$ and $f_{V_{\site,c}|H_{\site,c}}(v)$. To do so, we shall use the following lemma that gives the PDF of a transformed RV by a bijective function.

\begin{lem}
	Let $X$ be a RV and $f_X$ its PDF defined on an interval [a,b]. Let $Y$ be a RV such that $Y=g(X)$ with $g$ is a bijective function. The PDF of $Y$ $f_Y$ is given by 
	\begin{equation}
	f_Y(y)=f_X(g^{-1}(Y)) \left|\frac{d g^{-1}(y)}{dy} \right| \mathds{1}_{[g(a)\wedge g(b),g(a)\vee g(b)]}(y)
	\end{equation}
	where $g^{-1}$ is the inverse function of $g$, $g(a)\wedge g(b)=min(g(a),g(b))$ and $g(a)\vee g(b)=max(g(a),g(b))$.
	\label{transformRV}
\end{lem}   

The proof of this lemma is not provided here. To obtain the result, one can calculate the PDF by distinguishing between the case when $g$ is increasing and the case when it is decreasing.\\ 

Using lemma 1, the PDF $f_{\alpha_{\site,c}}$ of the RV $\alpha_{\site,c}$ is given by 

\begin{equation}
f_{\alpha_{\site,c}}(\alpha)=\frac{3}{2 \pi}  \mathds{1}_{[ \psi(m,\site)-\frac{2\pi}{3}c, \psi(m,\site)+\frac{2\pi}{3}(1-c)]}(\alpha).
\label{densityalpha}
\end{equation}

Similarly, from equation (\ref{phi1}), we can see that $\phi_{\site,c}$ is the transformation of the uniform RV $r_{\site,c}$ by the function $g(x)= atan(\frac{l_b}{|m-\site|})- atan(\frac{l_b}{x})$. Also, this RV depends on  $\alpha_{\site,c}$. One can easily verify that $g$ is strictly increasing on the interval $[0,\frac{2\delta}{3}U(\theta_{\site,c}-\vartheta_c)]$ and its inverse function $g^{-1}$  is given by
\begin{equation}
g^{-1}(y)=\frac{l_b}{tan(atan(\frac{l_b}{|m-\site|})-y)}.
\end{equation}

Hence, the PDF of $\phi_{\site,c}$ conditionally on $\alpha_{\site,c}$, denoted by $f_{\phi_{\site,c}|\alpha_{\site,c}}(\phi)$, is given by
{
	\begin{align}
	f_{\phi_{\site,c}|\alpha_{\site,c}}(\phi)&=\frac{3}{2\delta U(\theta_{\site,c}-\vartheta_c)}\Big(1+ \frac{1}{\tan^2(\phi-atan(\frac{l_b}{|m-\site|}))}\Big) \nonumber\\
	&\mathds{1}_{[atan(\frac{-|m-\site|}{l_b}),atan(\frac{2 \delta l_b U(\theta_{\site,c}-\vartheta_c)-3l_b |m-\site|}{2\delta|m-\site|U(\theta_{\site,c}-\vartheta_c)+3l_b^2})]}(\phi)
	\label{phidensity}
	\end{align}   
}  

Now, to characterize $G_{\site}(m)$, we apply once again Lemma 1 to $\alpha_{\site,c}$ and $\phi_{\site,c}$ transformed by the function $g$ such that $g=H$ for $\alpha_{\site,c}$ and $g=V$ for $\phi_{\site,c}$. Thence, using lemma 1 and equations (\ref{densityalpha}) and (\ref{phidensity}) we obtain   

\begin{align}
f_{H_{\site,c}}(h)&=\frac{-3h^{\frac{-1-2w_h}{2w_h}}}{4\pi w_h \sqrt{1-h^{\frac{-1}{w_h}}}} \mathds{1}_{[h_1\wedge h_2,h_1\vee h_2]}(h),  
\label{densityhorizental}
\end{align}  
with $h_1=H(arg(m-\site)-\frac{2\pi}{3}c)$ and $h_2=H(arg(m-\site)+\frac{2\pi}{3}(1-c))$. And

 \begin{align}
f_{V_{\site,c}|H_{\site,c}}(v)&=\Big(1+ \frac{1}{\tan^2(acos(v^{\frac{-1}{2w_v}})-atan(\frac{l_b}{|m-\site|}))}\Big)    \times \nonumber \\
& \frac{-3v^{\frac{-1-2w_v}{2w_v}}}{4w_v \delta U(\theta_{\site,c}-\vartheta_c)\sqrt{1-v^{\frac{-1}{w_v}}}} \mathds{1}_{[v_1\wedge v_2,v_1\vee v_2]}(v),
\label{densityvertical}
\end{align}  
with $v_1=V(atan(\frac{-|m-\site|}{l_b}))$ and $v_2=V(atan(\frac{2 \delta l_b U(\theta_{\site,c}-\vartheta_c)-3l_b |m-\site|}{2\delta|m-\site|U(\theta_{\site,c}-\vartheta_c)+3l_b^2}))$.\\

  Moreover, the joint PDF of $H(\alpha_{\site,c})$ and $V(\phi_{\site,c})$ denoted by $f_{H_{\site,c},V_{\site,c}}$ is given by
 
  \begin{equation}
  f_{H_{\site,c},V_{\site,c}}(h,v)=f_{V_{\site,c}|H_{\site,c}}(v) f_{H_{\site,c}}(h).
  \end{equation}
  
  To calculate the PDF of $Z_{h,v}=H(\alpha_{\site,c})V(\phi_{\site,c})$, one can use the Mellin transform of the distribution of $Z_{h,v}$, denoted by $f_{Z_{h,v}}$, defined by
  
  \begin{align}
  \mathcal{M}f_{Z_{h,v}}(s)&=\mathbb{E}[Z_{h,v}^{s-1}]=\int\int h^{s-1}v^{s-1}f_{H_{\site,c},V_{\site,c}}(h,v) dh dv \nonumber \\
  &=\int h^{s-1}\big( \int v^{s-1}f_{V_{\site,c}|H_{\site,c}}(v) dv \big) f_{H_{\site,c}}(h) dh
  \label{mellin}
  \end{align}
 Then, the PDF of the product of the horizontal and the vertical patterns $f_{Z_{h,v}}$ can be obtained by using the inverse Mellin transform as follows
  \begin{equation}
  f_{Z_{h,v}}(z)=\frac{1}{2\pi i}\int z^{-s}\mathcal{M}f_{Z_{h,v}}(s)ds.
  \end{equation} 
  
 Finally the distribution of $G_{\site}(m)$ can be obtained by the convolution of $ f_{Z_{h,v}}$ for $c \in \{1,2,3\}$.\\
 
 The mathematical expectation of $G_{\site}(m)$ is defined by
 
 \begin{equation}
 \mathbb{E}[Gs(m)]=\eta \sum_{c=1}^{3} \mathbb{E}[H(\alpha_{\site,c})V(\phi_{\site,c})],
 \end{equation}
  where $\mathbb{E}[H(\alpha_{\site,c})V(\phi_{\site,c})]$ is calculated by taking $s=2$ in equation (\ref{mellin}).\\
  
 When the downtilt is taken constant, the vertical antenna radiation pattern is no more a RV. Thus, the explicit expression of $\mathbb{E}[Gs(m)]$ conditionally on $m$ can be written as
  
  \begin{align}
  \mathbb{E}[Gs(m)]&=\frac{3\eta V(\phi_{\site,c})}{2\pi} \sum_{c=1}^{3}\int_{\upsilon_c-\frac{\pi}{3}}^{\upsilon_c+\frac{\pi}{3}} \cos^{-2w_h}(\theta - \psi(m,\site))d\theta \nonumber \\
  &=\frac{3\eta V(\phi_{\site,c})}{\pi} \int_{0}^{\pi}\cos^{-2w_h}(\theta) d\theta.
  \label{hhh1}
  \end{align}
  
 Equation (\ref{hhh1}) comes from the $2\pi-$periodicity of the function $\cos$ and Chasles's formula.\\

 Finally, by using $\int_{0}^{\frac{\pi}{2}} cos^z(x)=\frac{\sqrt{\pi}\Gamma(\frac{z+1}{2})}{2\Gamma(\frac{z}{2}+1)}$ for all complex number $z$ such that $\Re(z)>-1$, the explicit expression of the mathematical expectation of $G_{\site}(m)$ is given by
  
  \begin{equation}
  \mathbb{E}[Gs(m)]=\frac{3\eta V(\phi_{\site,c}) \Gamma(\frac{1}{2}-w_h)}{\sqrt{\pi}\Gamma(1-w_h)}
  \end{equation} 
  with $\Gamma(.)$ is the Euler Gamma function.

  \section{Interference characterization (1)}

  A mobile located at a position $m$ in the first sector ($c=1$) of the serving site $\site_{0}$ receives interference from the co-sectors and also from the other interfering sites in $\Lambda^*$. We define the individual $ISR$, denoted by $\mathcal{I}_{\site}(m)$, as the received power from an interfering site $\site$ divided by the useful power received by $m$ from the serving cell. Its expression is given by
  {
  	\begin{align}
  	\mathcal{I}_{\site}(m)&= \frac{P_r(m,\site)}{P_r(m,\site_{0})} \nonumber \\
  	&=r^{2b}|\site-m|^{-2b}G_{\site}(m)\tilde{\chi_{\site}},
  	\end{align}   
  }
  where $\tilde{\chi_{\site}}=10^{\frac{\tilde{Y}_{\site}}{10}}$ is a log-normal RV representing the ratio of the shadowing effect from interfering sites and the shadowing effect from the serving site (The ratio of two log-normal RVs is a log-normal RV), with $\tilde{Y}_{\site}$ is a Normal RV with mean 0 and variance $\tilde{\sigma}^2$. Also, we define the cumulative $ISR$ from all the interfering sites including the two other sectors of the serving site $\site_{0}$ as the sum over $\site \in \Lambda$ of all the individual ISRs. It is expressed as
  
  \begin{equation}
  \mathcal{I}(m)= -1+\sum_{\site \in \Lambda} r^{2b}|\site-m|^{-2b}G_{\site}(m) 10^{\frac{\tilde{Y}_{\site}}{10}}.
  \label{ISRtotal}
  \end{equation}

   $\mathcal{I}(m)$ is an infinite sum of independent positive RVs not identically distributed. Hence according to Theorem 9.2.a of \cite{jacod2012probability},
   
   \begin{equation}
   \mathbb{E}[ \mathcal{I}(m)]= -1+ \mathbb{E}[10^{\frac{\tilde{Y}_{\site}}{10}}] \sum_{\site \in \Lambda} r^{2b}|\site-m|^{-2b}\mathbb{E}[G_{\site}(m)].\nonumber
   \label{expec1}
   \end{equation}  
   
   In \cite{nasri2016analytical}, it has been shown that $\sum_{\site \in \Lambda^*} r^{2b}|\site-m|^{-2b}$ is a convergent series on $x=\frac{r}{\delta}$ that can be approximated as follows 
  
  \begin{equation}
  \sum_{\site \in \Lambda^*} r^{2b}|\site-m|^{-2b}\approx \frac {6x^{2b}}{\Gamma(b)^{2}}\sum_{h=0}^{+\infty}\frac{\Gamma(b+h)^{2}}{\Gamma(h+1)^{2}}\omega(b+h)x^{2h} 
  \label{ridha}
\end{equation}
\vspace*{0.3cm}
  \begin{equation}
  \text{ where }\omega(z) = 3^{-z}\zeta(z)\left( \zeta(z,\frac{1}{3})-\zeta(z,\frac{2}{3})\right), \nonumber
  \label{omega}
  \end{equation}
  with $\zeta(.)$ and $\zeta(.,.)$ are respectively the Riemann Zeta and Hurwitz Riemann Zeta functions \cite{abramowitz1964handbook}.\\
  
  Since $G_{\site}(m)\leq 1$, we have $\mathbb{E}(G_{\site}(m))< 1 $. By using (\ref{ridha}), we can easily prove that $\mathbb{E}[ \mathcal{I}(m)] < \infty$. Hence, according to Theorem 9.2.b of \cite{jacod2012probability}, $\mathcal{I}(m)$ converges almost surely.\\
  
   Unfortunately, the central limit theorem (CLT) can not be applied in this case. Also, the generalized CLT known as Lyapunov CLT and Lindeberg CLT conditions are not verified here. Hence, $\mathcal{I}(m)$ can not be approximated by a Gaussian RV.\\ 
  
%
  
  Additionally, we define the DL coverage probability $\varPi$ as the probability that a mobile user $m$ is able to achieve a threshold $SINR$ , denoted by $\gamma$, during the DL transmission
  
  \begin{equation}
  \varPi(\gamma)=\mathbb{P}(\Theta(m)>\gamma),
  \end{equation} 
  where $\Theta(m)$ is the $SINR$ of $m$ and it can be expressed in terms of the cumulative $ISR$ $\mathcal{I}(m)$ as follows
  
  \begin{equation}
  \Theta(m)=\frac{1}{\mathcal{I}(m) + y_0},
  \label{sinr}
  \end{equation}
  with $y_0=\frac{aN r^{2b}}{A P \chi_0 }$ and $N$ is the thermal noise power.\\
  
  Finally, the throughput is calculated using the upper bound of the well known Shannon's formula for a MIMO system $Tx \times Rx$, with $Tx$ and $Rx$ are respectively the number of transmit and receive antennas. Hence the throughput $C(m)$ of a user located at a position $m$ can be written as
  
  \begin{equation}
  C(m)=(Tx\wedge Rx)B_w log_2(1+\Theta(m)),
  \label{throughput}
  \end{equation} 
  with $B_w$ is the system bandwidth.
  
%

  \section{Numerical results and discussion}

  For numerical purpose, we consider 5 rings of interfering sites with an inter site distance $\delta=0.750km$. All BSs  are assumed to have the same height $l_b=30m$, transmit with a power level $P=40dBm$ and operate in a bandwidth $B_w$ of $20MHz$. The antenna gain is set to be $A=17dBi$. Also, the downlink thermal noise power is calculated for $B_w$ and set to $N=-93dBm$. The path loss exponent is considered to be $2b=3.5$, the propagation factor of an outdoor environment is $a=130dB$ and the standard deviation of the log-normal shadowing is $\sigma=5.5dB$. Finally, we assume that we have $2Rx$ antennas in user's terminals, the number of $Tx$ antennas depends on the chosen half power beam width ($\theta_{h3dB}$ and $\theta_{v3dB}$). This number with beamforming is greater than 2 and thus, the number of possible transmission layers is at most 2.\\

   We simulate the 3D beamforming model in MATLAB considering four values of $\theta_{h3dB}$ ($30\degree$, $20\degree$, $14\degree$ and $8\degree$) and a vertical half power beam width  $\theta_{v3dB}=8\degree$. We compare it to a simulated hexagonal tri-sectorized network without beamforming mechanisms with $\theta_{h3dB}=65$, $\theta_{v3dB}=32$ and a fixed downtilt angle $8\degree$.\\
   
\begin{figure}[tb]
  	\centering
  	\includegraphics[width=9cm,height=6cm]{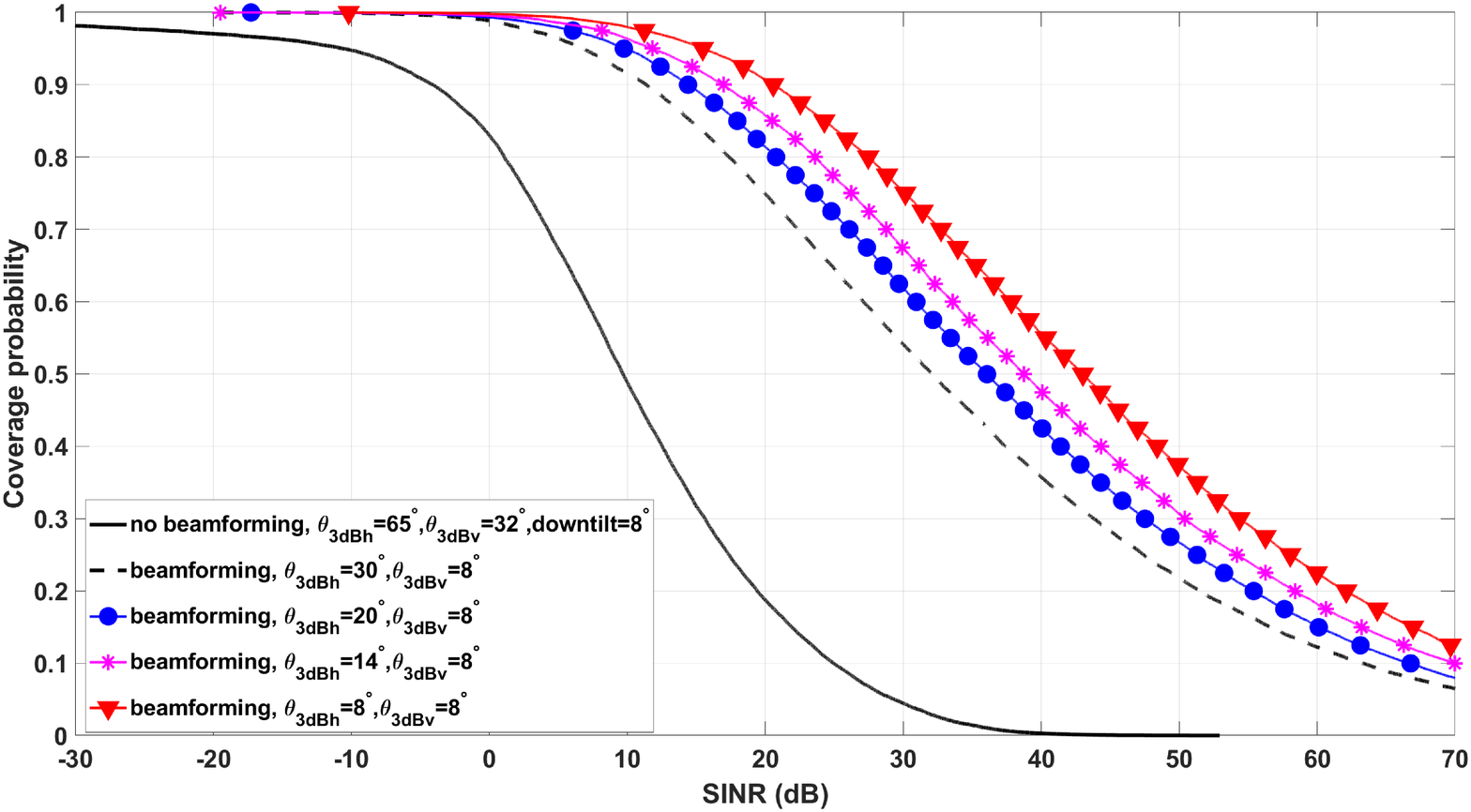}
  	\caption{Coverage probability: 3D beamforming impact.}
  	\label{beamwidthimpact}
  \end{figure}

    We plot in Fig. \ref{beamwidthimpact} the empirical coverage probability (CCDF of $SINR$) curves obtained by using Monte Carlo simulations for $20000$ mobile locations $m$. As we can see, 3D beamforming enhances significantly performance. For instance, with an $SINR$ threshold of $10dB$, the coverage probability increases from $66\%$ to $99\%$ when 3D beamforming is deployed with a horizontal beam width of $\theta_{h3dB}=8\degree$. Moreover, it can be observed that the coverage probability increases as the beam width decreases. Actually, the beam width is related to the number of transmit antennas used by BSs. When this number increases, the signal is focused on a specific zone of the cell. Hence, interference coming from neighboring sites are reduced significantly as the beam width decreases and the number of transmit antennas increases. This leads to an enhancement of $SINR$ and thus an enhancement of the coverage probability.\\

   Similarly in Fig. \ref{throughput}, we plot the user throughput as a function of the average load over interfering sites considering the 3D beamforming with two values of $\theta_{h3dB}$ ($16\degree$ and $8\degree$) and $\theta_{v3dB}=8\degree$. We compare results with the case of tri-sectorized hexagonal network without beamforming. As we can observe, with a tri-sectorized hexagonal network without 3D beamforming, the throughput is more sensitive to the average load variations over interfering sites and it decreases by almost $30\%$ when the average load increases from $1\%$ to $100\%$. With 3D beamforming, one can notice that the throughput increases significantly compared to the case without beamfroming. Also, it increases as the beam width decreases, which is in agreement with results of Fig. \ref{beamwidthimpact}. Additionally, we can see that the sensitivity to the average load and interference decreases in the case of 3D beamforming, especially with small half power beam widths i.e., when the number of transmit antennas increases.\\

\begin{figure}[tb]
 	\centering
 	\includegraphics[width=9cm,height=6cm]{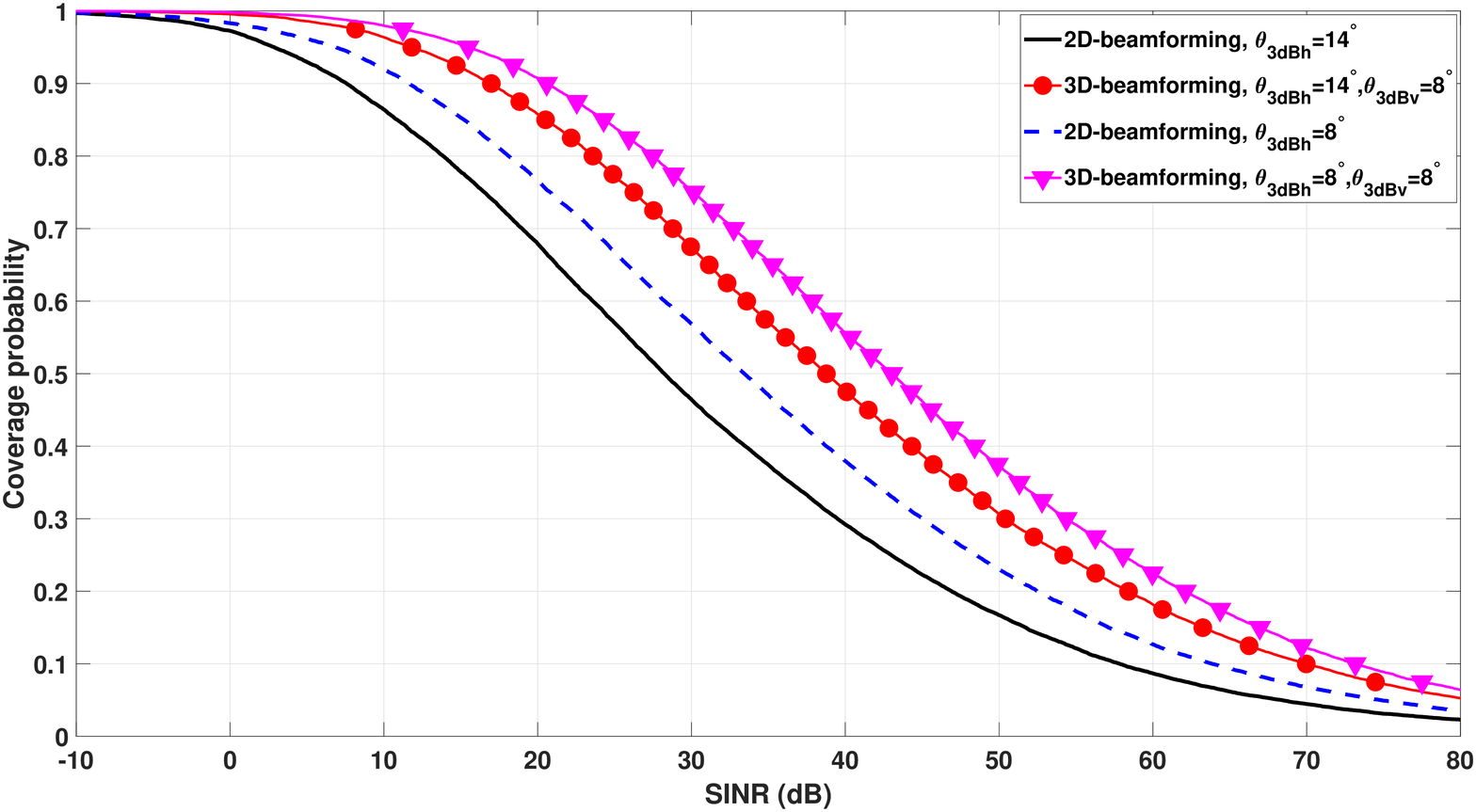}
 	\caption{Coverage probability comparison: 3D beamforming vs 2D beamforming.}
 	\label{2dvs3d}
 \end{figure}

   Fig. \ref{2dvs3d} shows a comparison between the coverage probability of a network adopting the 3D beamforming according to our model and the one of a network using a 2D beamforming where only a random horizontal radiation pattern is considered. The vertical component is taken constant and included in the antenna gain. Once again, we can notice that a small beam width leads to an increase of the coverage probability for the both models. Moreover, it is obvious that performance in terms of $SINR$ with 3D beamforming are better than 2D beamforming. Actually, most BSs use a linearly arranged array of antennas placed at the top of BSs. Unfortunately, the number of antennas can not be increased because of size constraints. Hence the interest of FD-MIMO, based on a 2D array of antennas, that offers the possibility to increase the number of transmit antennas and gives extra degrees of freedom in order to improve significantly the wireless system performance. Also, it provides the capability to adapt dynamically beam patterns in the horizontal and vertical dimensions.\\

    \begin{figure}[tb]
    	\centering
    	\includegraphics[width=9cm,height=6cm]{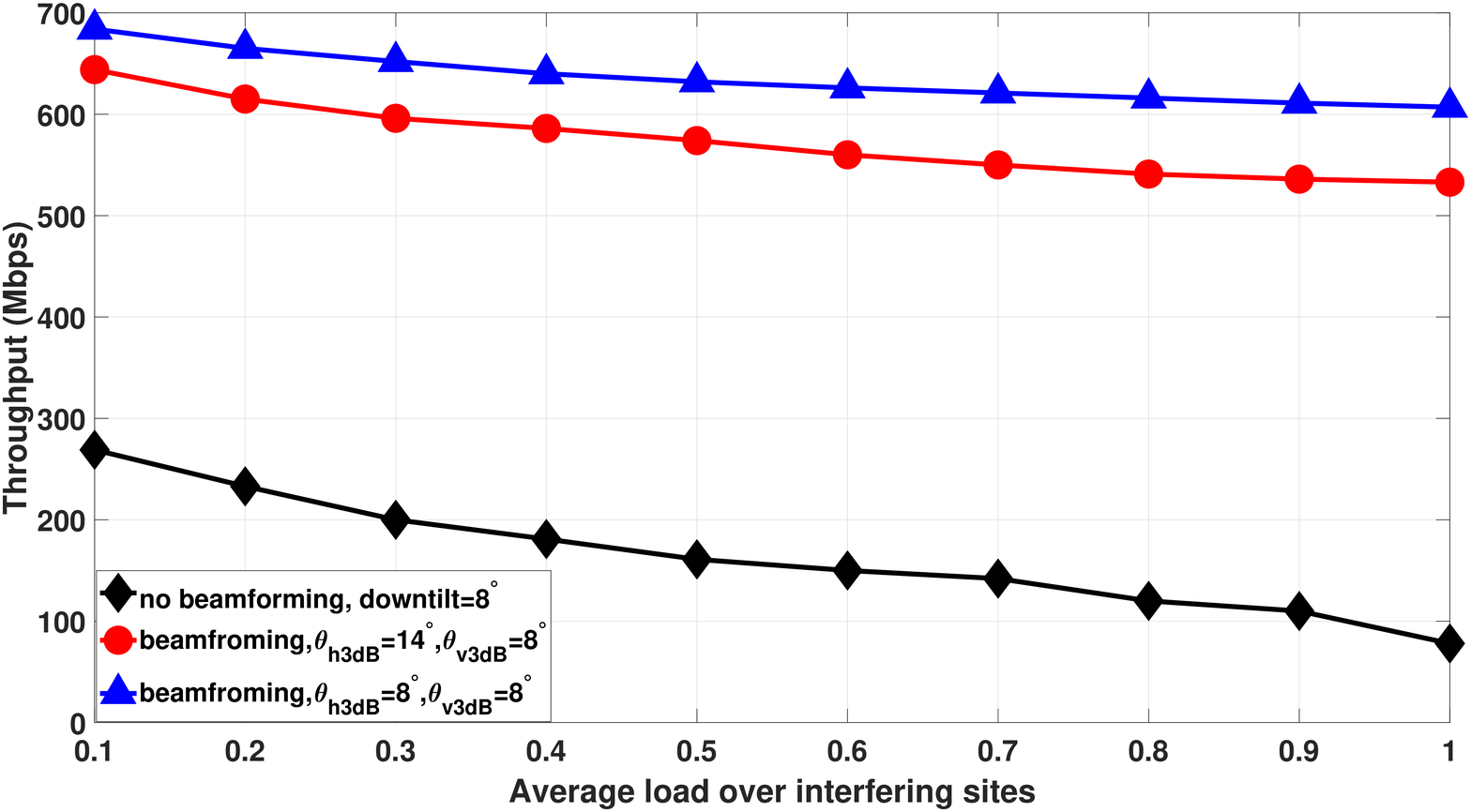}
    	\caption{Throughput variation with the average load over interfering sites: comparison between 3D beamforming model and hexagonal tri-sectorized network.}
    	\label{throughput}
    \end{figure}

\begin{figure}[tb]
	\centering
	\includegraphics[width=9cm,height=6cm]{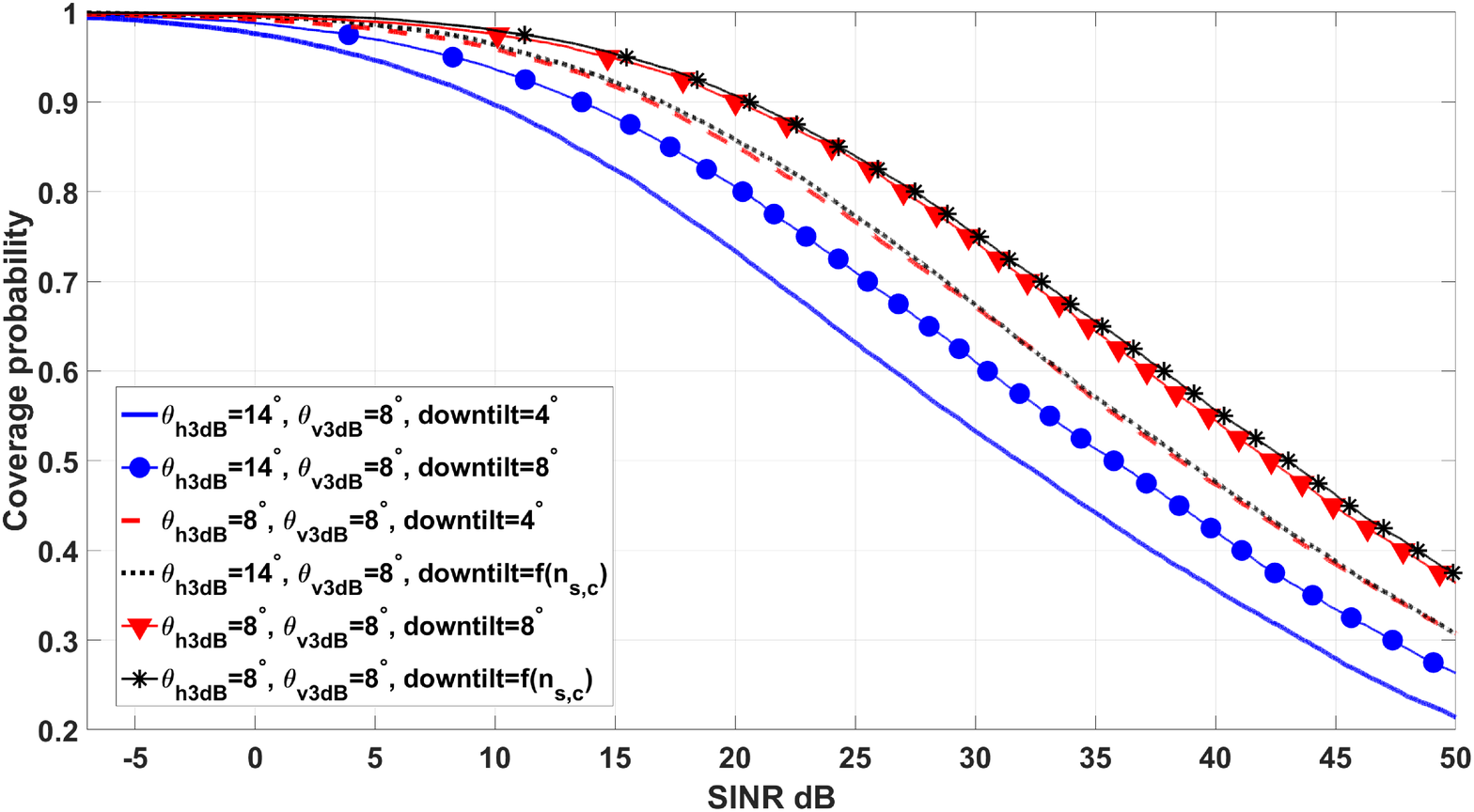}
	\caption{Downtilt $\tilde{\phi}_{\site,c}$ impact: variable vs fixed downtilt.}
	\label{tiltimpact}
\end{figure}

Finally in Fig. \ref{tiltimpact}, we plot the coverage probability with the 3D beamforming model, presented in this paper, for 2 values of $\theta_{h3dB}$ ($8\degree$ and $14\degree$) and $\theta_{v3dB}=8\degree$. We compare it with the coverage probability of the same network, but instead of taking the downtilt angle $\tilde{\phi}_{\site,c}$ as a function of the mobile location $n_{\site,c}$, we consider two values of this angle, $\tilde{\phi}_{\site,c}=4\degree$ and $\tilde{\phi}_{\site,c}=8\degree$, close to the values used in practical systems. One can observe that when the horizontal beam width is set to $14\degree$, the 3D beamforming model with variable downtilt angle has better performance than the case of fixed downtilt for the two values of $\tilde{\phi}_{\site,c}$. However, with a horizontal beam width of $8\degree$, we can notice that the coverage probability when $\tilde{\phi}_{\site,c}=8\degree$ is getting closer to the one of the network with 3D beamforming. Hence, using a large number of transmit antennas to concentrate the beam to a specific location, especially in the horizontal dimension, leads to a significant enhancement of performances. Moreover, for the vertical dimension, it has been shown in several papers that there exists an optimal downtilt angle for which the performance are enhanced significantly. In practical systems, we take always a downtilt close to 6$\degree$.

\section{Conclusions}

In this paper, we have proposed an analytical model for 3D beamforming where antenna radiation patterns depend on the spatial distribution of users' locations in the plane. We have shown, through system level simulations, that  the 3D beamforming reduces significantly interference and thus enhances the $SINR$ and users throughput in downlink. A comparison between the proposed 3D model and the traditional 2D beamforming, where only the azimuthal plane is considered, shows better performance with 3D beamforming. Hence the importance of FD-MIMO, where a large number of antennas are arranged in a 2D array, that makes possible to adapt beamforming also in the vertical dimension. A further extension of this work can include some analytical results regarding the distribution of beamforming radiation and the derivation of an upper bound of the coverage probability using concentration inequalities. Also, one can study the imperfection of beam steering especially for indoor users by considering the same assumptions as this work.  
  
\balance

 \bibliographystyle{IEEEtran}
 \bibliography{IEEEabrv,TelecomReferences}

\begin{thebibliography}{10}
\providecommand{\url}[1]{#1}
\csname url@samestyle\endcsname
\providecommand{\newblock}{\relax}
\providecommand{\bibinfo}[2]{#2}
\providecommand{\BIBentrySTDinterwordspacing}{\spaceskip=0pt\relax}
\providecommand{\BIBentryALTinterwordstretchfactor}{4}
\providecommand{\BIBentryALTinterwordspacing}{\spaceskip=\fontdimen2\font plus
\BIBentryALTinterwordstretchfactor\fontdimen3\font minus
  \fontdimen4\font\relax}
\providecommand{\BIBforeignlanguage}[2]{{%
\expandafter\ifx\csname l@#1\endcsname\relax
\typeout{** WARNING: IEEEtran.bst: No hyphenation pattern has been}%
\typeout{** loaded for the language `#1'. Using the pattern for}%
\typeout{** the default language instead.}%
\else
\language=\csname l@#1\endcsname
\fi
#2}}
\providecommand{\BIBdecl}{\relax}
\BIBdecl

\bibitem{gandhi2014significant}
A.~D. Gandhi, ``Significant {G}ains in {C}overage and {D}ownlink {C}apacity
  {F}rom {O}ptimal {A}ntenna {D}owntilt for {C}losely-{S}paced cells in
  wireless networks,'' in \emph{Wireless and Optical Communication Conference
  (WOCC), 2014 23rd}.\hskip 1em plus 0.5em minus 0.4em\relax IEEE, 2014, pp.
  1--6.

\bibitem{bai2015coverage}
T.~Bai and R.~W. Heath, ``Coverage and {R}ate {A}nalysis for
  {M}illimeter-{W}ave {C}ellular {N}etworks,'' \emph{IEEE Transactions on
  Wireless Communications}, vol.~14, no.~2, pp. 1100--1114, 2015.

\bibitem{koppenborg20123d}
J.~Koppenborg, H.~Halbauer, S.~Saur, and C.~Hoek, ``3d {B}eamforming {T}rials
  {W}ith an {A}ctive {A}ntenna {A}rray,'' in \emph{Smart Antennas (WSA), 2012
  International ITG Workshop on}.\hskip 1em plus 0.5em minus 0.4em\relax IEEE,
  2012, pp. 110--114.

\bibitem{halbauer2012interference}
H.~Halbauer, S.~Saur, J.~Koppenborg, and C.~Hoek, ``Interference {A}voidance
  {W}ith {D}ynamic {V}ertical {B}eamsteering in {R}eal {D}eployments,'' in
  \emph{Wireless Communications and Networking Conference Workshops (WCNCW),
  2012 IEEE}.\hskip 1em plus 0.5em minus 0.4em\relax IEEE, 2012, pp. 294--299.

\bibitem{yang2018optimal}
J.~Yang, M.~Ding, G.~Mao, Z.~Lin, D.-g. Zhang, and T.~H. Luan, ``Optimal {B}ase
  {S}tation {A}ntenna {D}owntilt in {D}ownlink {C}ellular {N}etworks,''
  \emph{arXiv preprint arXiv:1802.07479}, 2018.

\bibitem{seifi2012impact}
N.~Seifi, M.~Coldrey, M.~Matthaiou, and M.~Viberg, ``Impact of {B}ase {S}tation
  {A}ntenna {T}ilt on {T}he {P}erformance of {N}etwork {MIMO} {S}ystems,'' in
  \emph{IEEE 75th Vehicular Technology Conference, VTC Spring 2012, Yokohama, 6
  May-9 June 2012}, 2012.

\bibitem{nasri2016analytical}
R.~Nasri and A.~Jaziri, ``{A}nalytical {T}ractability of {H}exagonal {N}etwork
  {M}odel {W}ith {R}andom {U}ser {L}ocation,'' \emph{IEEE Transactions on
  Wireless Communications}, vol.~15, no.~5, pp. 3768--3780, 2016.

\bibitem{rachad2018interference}
J.~Rachad, R.~Nasri, and L.~Decreusefond, ``Interference analysis in dynamic
  tdd system combined or not with cell clustering scheme,'' in \emph{2018 IEEE
  87th Vehicular Technology Conference (VTC Spring)}.\hskip 1em plus 0.5em
  minus 0.4em\relax IEEE, 2018, pp. 1--5.

\bibitem{andrews2011tractable}
J.~G. Andrews, F.~Baccelli, and R.~K. Ganti, ``A {T}ractable {A}pproach to
  {C}overage and {R}ate in {C}ellular {N}etworks,'' \emph{IEEE Transactions on
  communications}, vol.~59, no.~11, pp. 3122--3134, 2011.

\bibitem{blaszczyszyn2016spatial}
B.~B{\l}aszczyszyn and M.~K. Karray, ``Spatial {D}istribution of {T}he
  {S}{I}{N}{R} in {P}oisson {C}ellular {N}etworks {W}ith {S}ector {A}ntennas,''
  \emph{IEEE Transactions on Wireless Communications}, vol.~15, no.~1, pp.
  581--593, 2016.

\bibitem{mogensen1997preliminary}
P.~E. Mogensen, K.~I. Pedersen, P.~Leth-Espensen, B.~Fleury, F.~Frederiksen,
  K.~Olesen, and S.~L. Larsen, ``Preliminary measurement results from an
  adaptive antenna array testbed for gsm/umts,'' in \emph{1997 IEEE 47th
  Vehicular Technology Conference. Technology in Motion}, vol.~3.\hskip 1em
  plus 0.5em minus 0.4em\relax IEEE, 1997, pp. 1592--1596.

\bibitem{mogensen2007lte}
P.~Mogensen, W.~Na, I.~Z. Kov{\'a}cs, F.~Frederiksen, A.~Pokhariyal, K.~I.
  Pedersen, T.~Kolding, K.~Hugl, and M.~Kuusela, ``{LTE} {C}apacity {C}ompared
  to {T}he {S}hannon {B}ound,'' in \emph{Vehicular Technology Conference, 2007.
  VTC2007-Spring. IEEE 65th}.\hskip 1em plus 0.5em minus 0.4em\relax IEEE,
  2007, pp. 1234--1238.

\bibitem{jacod2012probability}
J.~Jacod and P.~Protter, \emph{Probability {E}ssentials}.\hskip 1em plus 0.5em
  minus 0.4em\relax Springer Science \& Business Media, 2012.

\bibitem{abramowitz1964handbook}
M.~Abramowitz and I.~A. Stegun, \emph{Handbook of {M}athematical {F}unctions:
  with formulas, graphs, and mathematical tables}.\hskip 1em plus 0.5em minus
  0.4em\relax Courier Corporation, 1964, no.~55.

\end{thebibliography}

\end{document}